\documentclass[prc,aps,superscriptaddress,showpacs,nofootinbib,twocolumn]{revtex4-1}
\usepackage{amssymb}
\usepackage[dvips]{graphicx}
\usepackage[english]{babel}
\usepackage{indentfirst}
\usepackage{amsxtra}
\usepackage{amsmath}
\usepackage{supertabular}
\usepackage{multirow}
\usepackage[mathcal]{eucal}
\usepackage[normalem]{ulem}
\usepackage[usenames]{color}
\usepackage{ulem}

\begin{document}
\title {\textbf{Quenching of Gamow-Teller strengths and two particle - two hole configurations}}
\author{Danilo Gambacurta}
\affiliation{INFN-LNS, Laboratori Nazionali del Sud, 95123 Catania, Italy}
\author{Marcella Grasso}
\affiliation{Universit\'e Paris-Saclay, CNRS/IN2P3, IJCLab, 91405 Orsay, France}

\begin{abstract}
 
We apply the charge-exchange subtracted second random-phase approximation (SSRPA), based on Skyrme functionals, to investigate Gamow-Teller resonances in several closed-shell and closed-subshell nuclei, located in different regions of the nuclear chart. 

After having discussed the SSRPA findings obtained within different approximation schemes in $^{48}$Ca, we compare our results with {\it{ab-initio}} coupled-cluster predictions available for C and O isotopes, where two-body currents are included. Our integrated strenghts, 
obtained by using one-body transition operators, are lower compared to the corresponding {\it{ab-initio}} results. This indicates that, within our model, quenching effects are mainly driven by the inclusion of two particle - two hole configurations  and that the role of a two-body contribution in the transition operator is less important than in the coupled-cluster approach. 

By analyzing heavier nuclei, $^{90}$ Zr and $^{132}$Sn, we confirm the same conclusions that we have recently drawn for $^{48}$Ca: the inclusion of two particle - two hole configurations is very effective in our model for providing strengths which are significantly more quenched than in other theoretical models and, thus, in better agreement with the experimental measurements. This occurs because two particle - two hole configurations have a density which strongly increases with the excitation energy. Their inclusion thus pushes a significant amount of the strength to higher energies, compared to what happens in other theoretical models, reducing in this way the cumulative sum of the strength up to excitation energies around 20-30 MeV.  

\end{abstract}

\maketitle

\section{Introduction} \label{intro}
We recently analyzed Gamow-Teller (GT) strengths in $^{48}$Ca and $^{78}$Ni with the self--consistent charge--exchange (CE) SSRPA \cite{gamba2020}, based on Skyrme effective interactions \cite{sk1,sk2,vautherin}, in the framework of energy--density--functional (EDF) theories. 
A subtraction procedure, employed to handle instabilities, to avoid overcounted correlations, and to regularize ultraviolet divergences \cite{tse2013,gamba2015}, was fully applied by inverting the matrix $A_{22}$, which represents the two particle - two hole (2p2h) sector of the SSRPA matrix. Reference \cite{gamba2020} illustrates the first application of this procedure within a CE version of the SSRPA model.  

The agreement of the predicted GT strength  with the corresponding experimental distribution in $^{48}$Ca \cite{yako} was shown to be significantly improved, compared to  other available theoretical predictions (see for example Refs. \cite{cao,niu}). In these models the Ikeda sum rule \cite{ike}, integrated up to an excitation energy of 20-30 MeV, is systematically overestimated and is predicted much closer to the full value $3(N-Z)$ than the experimental measurement. When this overestimation occurs, it is customary to 
resort to {\it{ad-hoc}} quenching factors in the GT operator, which are extensively employed (both in mean-field and in beyond-mean-field approaches within EDF models) to improve the agreement with the experimental measurements.  
For the first time, in Ref. \cite{gamba2020}, the experimental quenching of the Ikeda sum rule could be well accounted for in $^{48}$Ca without adopting {\it{ad-hoc}} factors. 

This achievement might have implications for future applications of the CE-SSRPA model to the evaluation of the nuclear matrix elements (NMEs) entering in neutrinoless-double-beta ($0\nu \beta \beta$) decay half-lives, for example in $^{48}$Ca, which is expected to be the lightest $\beta \beta$ emitter.  
The mechanisms underlying GT resonances and $\beta$ decays have a deeply different nature. Nevertheless,
a GT-type term turns out to be the leading contribution in  $0\nu \beta \beta$-decay NMEs. In addition,   
it is known that most of the theoretical models which resort to {\it ad-hoc} quenched GT operators also tend to overestimate single $\beta$-decay rates (compared to available data) and need, in practice, to quench by hand the weak interaction axial-vector coupling constant $g_A$ by a factor of $\approx$ 0.75 (see for instance Refs. \cite{towner,wilkinson,brown,chou,martinez}). 
The impact that a quenched $g_A$ value would have in particular on the sensitivity of $0\nu\beta\beta$ measurements was analyzed in Ref. \cite{suhonen}  and was shown to be far from being negligible. 

For all these reasons, 
a prediction of $0\nu\beta\beta$ NMEs (based on the bare value of the coupling constant $g_A$) would be significantly more reliable if obtained with a model where, 
coherently, {\it{ad-hoc}} quenching procedures are not required to describe satisfactorily experimental GT 
strengths and single $\beta$-decay rates. We also mention that another direction explored in the literature consists in evaluating rigorously the quenching of the transition operator by a proper computation of the correlations missing in the model as done, for example, in the shell-model calculations of Ref. \cite{coraggio}, based on realistic  potentials and the many-body-perturbation-theory approach. 

The low-energy part of the GT strength of $^{78}$Ni was used in Ref. \cite{gamba2020} to evaluate the 
$\beta$-decay half-life in this nucleus (using the bare value of $g_A$). The comparison with the experimental value \cite{hosmer,david} was found to be quite good.

The success of the CE-SSRPA model was ascribed in Ref. \cite{gamba2020} to the explicit inclusion of 2p2h configurations. As a matter of fact, these configurations 
have a density which increases with the excitation energy, producing a high--energy tail in the spectrum as was anticipated by some authors in the literature (see for instance Refs. \cite{bertsch,brink,arima}) but never demonstrated in practice before the work published in Ref. \cite{gamba2020}. Such a mechanism effectively pushes part of the strength towards higher excitation energies and leads to a better agreement with the reduced percentage of the Ikeda sum rule which is measured experimentally. What was called for many years ``the puzzle of the missing strength" could be finally elucidated. 

In the present work, we aim to extend our analysis of GT strengths based on the CE-SSRPA model to other nuclei, all closed-shell or closed-subshell, spanning different regions of the nuclear chart. 

The calculations performed for $^{48}$Ca in Ref.  \cite{gamba2020} were carried out by applying a full subtraction procedure, that is, without resorting to a diagonal approximation in the $A_{22}$ matrix to invert. We show in the present work that this is indeed a crucial point in GT calculations for $^{48}$Ca, where the coupling between 2p2h configurations has to be entirely taken into account not only in the matrix to diagonalize but also in the matrix to invert for the subtraction. 
Such fully implemented calculations require a considerable computational effort. 
For this reason, a special care must be devoted to the numerical implementation of the subtraction procedure and we employ here a method based on a 2$\times$2 block matrix inversion that can be found in standard textbooks \cite{inversion1,inversion2,inversion3}.This procedure allows us to strongly reduce the computational effort to achieve the inversion, making it feasible also for medium-mass nuclei.
 
After this preliminary analysis on $^{48}$Ca, we examine two lighter nuclei, $^{14}$C and $^{22}$O (where a full inversion of the matrix $A_{22}$ may be carried out much more easily), in order to compare our results with the {\it{ab-initio}} coupled-cluster predictions of Ref. \cite{ekstrom}. Such a comparison is particularly interesting because, in this specific {\it{ab-initio}} scheme, 2p2h configurations are also included. 
The authors of Ref. \cite{ekstrom} show in particular the importance of using, together with a a three-nucleon force, coherent two-body currents in the transition operator.
 
In our case, we employ effective Skyrme functionals. As far as three-body forces are concerned, these functionals contain effectively a three-body contribution through the density-dependent two-body term appearing in the interaction. On the other side, we systematically use one-body GT transition operators. 
Hence, the comparison with the  quenching of GT strengths predicted with the coupled-cluster approach will tell us whether, within our model, the inclusion of a two-body part in the transition operator is (or not) an important missing ingredient. If we find strengths which are similarly or more strongly quenched compared to those of Ref. \cite{ekstrom} for the same nuclei, we may conclude that the inclusion of a two-body transition operator is not  crucial  in our case.  

We then treat two nuclei heavier than $^{48}$Ca, namely $^{90}$Zr and $^{132}$Sn, and compare strengths and integrated strengths with the available experimental data \cite{zr1,zr2,sn}. We check in particular whether, also in these cases as for $^{48}$Ca, the experimental quenching of the integrated strengths is better reproduced compared to other theoretical models. 

The article is organized as follows. Section \ref{form} briefly provides details on the performed calculations and the used transition operator. 
In Sec. \ref{48ca}, we discuss the GT strength in
 $^{48}$Ca  and show that a full inversion of the $A_{22}$ matrix is necessary. 
 We present in Sec. \ref{light} results for the nuclei $^{14}$C and $^{22}$O as well as the comparisons with Ref. \cite{ekstrom}. Section \ref{heavy} illustrates results for the heavier nuclei $^{90}$Zr and  $^{132}$Sn and the comparisons with the corresponding experimental measurements, published in Refs. \cite{zr1,zr2} and  \cite{sn}, respectively. Conclusions are drawn in Sec. \ref{conclu}.

\section{Details on the calculations} \label{form}

The reader may find details on the CE second-random-phase-approximation (CE-SRPA) formalism (without subtraction) in Ref. \cite{brink} and details on the SSRPA formalism (for charge-conserving excitations) in Ref. \cite{gamba2015}. The first CE-SSRPA results (that is, CE results obtained by applying a subtraction procedure within the SRPA scheme) were presented in Ref. \cite{gamba2020}, as already mentioned in Sec. \ref{intro}. 

In this work, CE-SSRPA calculations are carried out on top of Hartree-Fock ground states, using the Skyrme parametrization SGII \cite{sgii1,sgii2}. 
The cutoff on one particle-one hole (1p1h) configurations is set at 60 MeV, whereas the one on 2p2h configurations is in most cases chosen equal to a value between 40 and 50 MeV. We have checked that all the used cutoff values ensure that the relevant physics is always included leading in fact to cutoff independent results. 

We employ GT one-body transition operators 

\begin{equation}
\hat{O}^{\pm}=\sum_{i=1}^{A} \sum_{\mu} \sigma_{\mu}(i) \tau^{\pm}(i),
\label{oper}
\end{equation} 
for a nucleus with $A$ nucleons; $\tau^{\pm}$ are the 
 isospin raising ($+$) and lowering ($-$) operators, $\tau^{\pm}=t_x \pm it_y$, and $\sigma_{\mu}$ is the spin operator. 

The $\hat{O}^+$ operator generates the GT$^+$ strength (a neutron is added and a proton is removed), whilst the $\hat{O}^-$  operator produces the GT$^-$ strength (a neutron is removed and a proton is added).  
The Ikeda GT sum rule \cite{ike}, relating the integrated strengths $S$ of the GT$^-$ and the GT$^+$ spectra to the number of neutrons $N$ and protons $Z$ of the nucleus, has the following expression: 
\begin{equation}
S_{GT^-}-S_{GT^+}=3(N-Z).
\label{ikeda}
\end{equation}
It is known that the strength $S_{GT^-}$ is dominant in nuclei having a neutron excess. 
This sum rule is model independent and can be easily deduced by using the properties of the isospin operators if the condition of completeness of states is fulfilled. 
It was underlined in particular in Ref. \cite{gamba2020} that, within the random-phase approcimation (RPA) and extended-RPA models, this sum rule holds if the quasiboson approximation is used. 
Since the inclusion of 2p2h configurations pushes a non negligible part of the SSRPA strength to high energies, the full value $3(N-Z)$ is reached only by integrating up to a very high excitation energies, much higher than in RPA. As already stressed, this mechanism allows for a strongly improved agreement of the integrated strength with the experimental cumulative Ikeda sum in the nucleus $^{48}$Ca.

\section{Diagonal approximation in $^{48}$Ca} \label{48ca}

\begin{figure}
\includegraphics[scale=0.32]{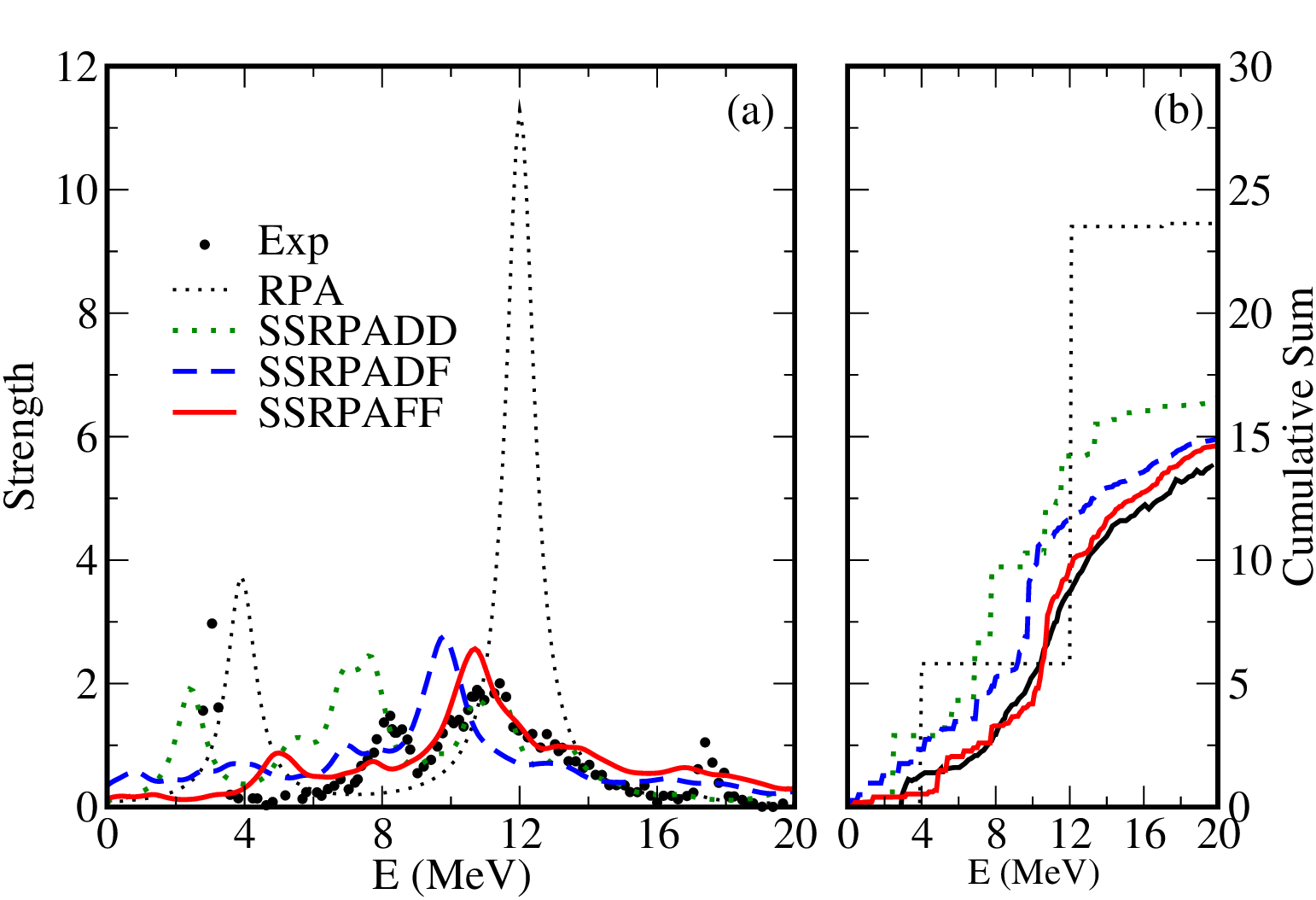}
\caption{(a) GT$^-$ strengths obtained with the Skyrme interaction SGII for the nucleus $^{48}$Ca in MeV$^{-1}$. Experimental data are extracted from Ref. \cite{yako}. RPA and all SSRPA strengths are obtained by folding with a Lorentzian having a width of 1 MeV. (b) Cumulative sum of the strength up to the excitation energy of 20 MeV. See  text for more details.}
\label{ffd}
\end{figure}

The full GT$^-$ spectrum for $^{48}$Ca was published in Ref. \cite{gamba2020}, where the SGII Skyrme interaction was used. The same interaction is employed here. This spectrum corresponds to the SSRPAFF curve shown in 
panel (a) of Fig. \ref{ffd}, where different GT$^-$ strengths are plotted. All the theoretical distributions are obtained by folding the discrete spectra with a Lorentzian having a width of 1 MeV. The experimental data are extracted from Ref. \cite{yako}. As indicated above, the spectrum obtained with a full calculation is denoted by the acronym SSRPAFF. The acronym SSRPADD refers to a calculation where the diagonal approximation is used twice for the matrix $A_{22}$, both in the inversion and in the diagonalization procedures. Finally, SSRPDF indicates a calculation where the diagonal approximation for the matrix $A_{22}$ is adopted only in the subtraction procedure, where this matrix is inverted, whereas $A_{22}$ is fully treated in the diagonalization. Also RPA results are shown for comparison. 
 
Figure \ref{ffd}  shows that the SSRPAFF model leads to a visible improvement of the results compared to RPA, as already discussed in Ref. \cite{gamba2020}. The double diagonal approximation (SSRPADD) provides a worse reproduction of the full spectrum, yet showing a considerable improvement with respect to RPA. The hybrid SSRPADF computation leads to better results, but still different from the SSRPAFF ones.  
Whereas Ref. \cite{gamba2015} showed for charge-conserving excitations that hybrid ``DF"-type results were not strongly different compared to the full spectrum, Fig. \ref{ffd} clearly indicates that this is no longer the case here.
Panel (b) of Fig.  \ref{ffd} confirms that the SSRPADF scheme reproduces sensibly less well than SSRPAFF the experimental strength and this is visible in the displayed cumulative sums. Such a result indicates that, in the calculations done for $^{48}$Ca, the inversion of the matrix in the subtraction procedure cannot be carried out by adopting a simple diagonal approximation in the matrix and needs a more accurate treatment.

\section{Lighter nuclei: $^{14}$C and $^{22}$O.} \label{light}

Before treating nuclei heavier than $^{48}$Ca, we address and check in lighter systems another aspect of our calculations. 

We have chosen the nuclei  $^{14}$C and $^{22}$O as illustrations, because the GT Ikeda sum rule was computed for these nuclei in Ref. \cite{ekstrom} with coupled-cluster calculations. 
In these calculations based on effective-field theories \cite{epelbaum}
the GT transition operator, Eq. (\ref{oper}), is corrected by the inclusion of two-body currents, in a coherent way with the inclusion of three-body forces. 
The impact of the inclusion of two-body currents in the transition operator was analyzed in Ref. \cite{ekstrom} by studying three nuclei, $^{14}$C, $^{22}$O, and $^{24}$O.  The authors found that such a modification of the operator induces a change in the total Ikeda sum rule which is reduced compared to the value $3(N-Z)$
and results equal to $3(N-Z)$ times a quenching factor $q^2 \sim 0.84 - 0.92$. The authors also checked that the value $3(N-Z)$ is obtained if the transition operator contains only a one-body part. Hence, they concluded that the inclusion of two-body currents may solve the 
discrepancy between theoretical GT integrated strengths and experimental measurements. 
They found that 70-80 \% of the total (reduced) strength was exhausted for the three nuclei, up to the excitation energy of 10 MeV. Similar conclusions were drawn in Ref. \cite{gysbers}, where 
the $\beta$-decay rate was computed for $^{100}$Sn within the same theoretical framework. 
 
\begin{figure}
\includegraphics[scale=0.32]{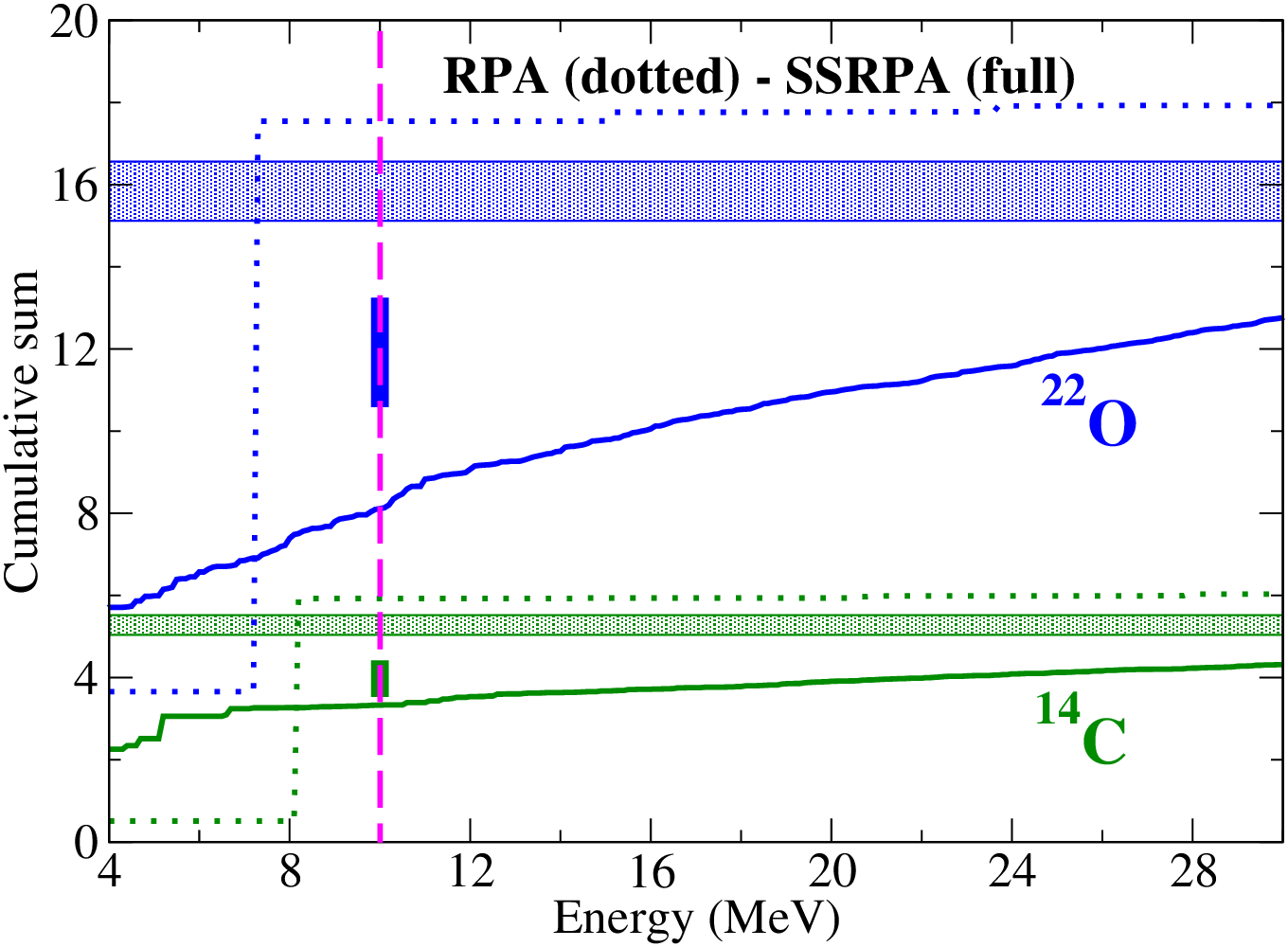}
\caption{RPA and SSRPA cumulative sums for the $S_{GT^-}$ strength for the nuclei $^{22}$O (blue) and $^{14}$C (green). The blue and green horizontal areas represent the reduction of the total Ikeda sum rule $S_{GT^-}-S_{GT^+}$ that was predicted for these nuclei in Ref. \cite{ekstrom}. The vertical dashed magenta line indicates 10 MeV of excitation energy. 
The blue and green vertical intervals there correspond to the predictions of Ref. \cite{ekstrom} for the amount (70-80 \%) of the total (reduced) sum rule exhausted up to 10 MeV for the two nuclei (see text). }
\label{ik}
\end{figure}

We show the results of Ref. \cite{ekstrom} in Fig. \ref{ik}, together with RPA and SSRPA predictions obtained with SGII, for the nuclei $^{14}$C and $^{22}$O, as illustrations. 
As for the case of $^{48}$Ca, we have checked that also for these lighter nuclei a hybrid calculation (diagonal approximation in the matrix to invert) reveals itself to be a rather poor approximation and a full calculation is necessary. 

The RPA and SSRPA cumulative sums for the $S_{GT^-}$ strength are plotted in the interval of excitation energies from 4 to 30 MeV. The excitation energy of 10 MeV is indicated by a vertical dashed magenta line. The $S_{GT^-}$ strength is dominant compared to the $S_{GT^+}$ one, as one can easily check in the figure, where the RPA cumulative sums almost attain the total Ikeda values of 18 and 6 for 
$^{22}$O and $^{14}$C, respectively. The two horizontal colored areas represent, for the two nuclei, the total Ikeda sum rule reduced by the $q^2$ value deduced in Ref. \cite{ekstrom}. The vertical blue and green intervals at 10 MeV correspond to 70-80 \% of each of these reduced sum rules. We may notice that, at 10 MeV, the RPA cumulative sums are already above these vertical intervals in both cases, whereas the SSRPA values are located below. 
The SSRPA values remain below the two colored horizontal areas in the whole interval of energies shown in the figure, up to 30 MeV. 
This indicates that the way the quenching is accounted for in our model (where the total sum rule $3(N-Z)$ is fulfilled) provides coherent results compared to those of Ref. \cite{ekstrom}, contrary to what happens in RPA, where the values are located above the two horizontal areas already at 8 MeV of excitation energy. 
This also means that, in practice, the inclusion of a two-body part in the transition operator does not seem to be an important ingredient for the description of the quenching in our model. We finally stress that we prefer to work within a model where the total sum rule $3(N-Z)$ is expected to be exhausted (even if at high excitation  energies), because this  provides a precise reference value and, thus, a robust check for the theoretical calculations.

\section{Heavier nuclei: $^{90}$Zr and $^{132}$Sn} \label{heavy}

The $^{90}$Zr(p,n)  and  $^{90}$Zr(n,p)  reactions were performed at the Research Center for Nuclear Physics and the results were published in 1997 \cite{zr1}
and in 2005 \cite{zr2}, respectively. A consistent study of data coming from both (p,n) and (n,p) channels led in Ref. \cite{zr2} to the estimation of the experimental quenching for the GT Ikeda sum rule. In particular, the GT$^-$ strength integrated up to 50 MeV was found to be equal to 29.3, very close to the full value of the Ikeda sum rule for this nucleus ($3(N-Z)=$ 30), whereas the GT$^+$ strength integrated up to the same excitation energy was found to be equal to 2.9.

\begin{figure}
\includegraphics[scale=0.32]{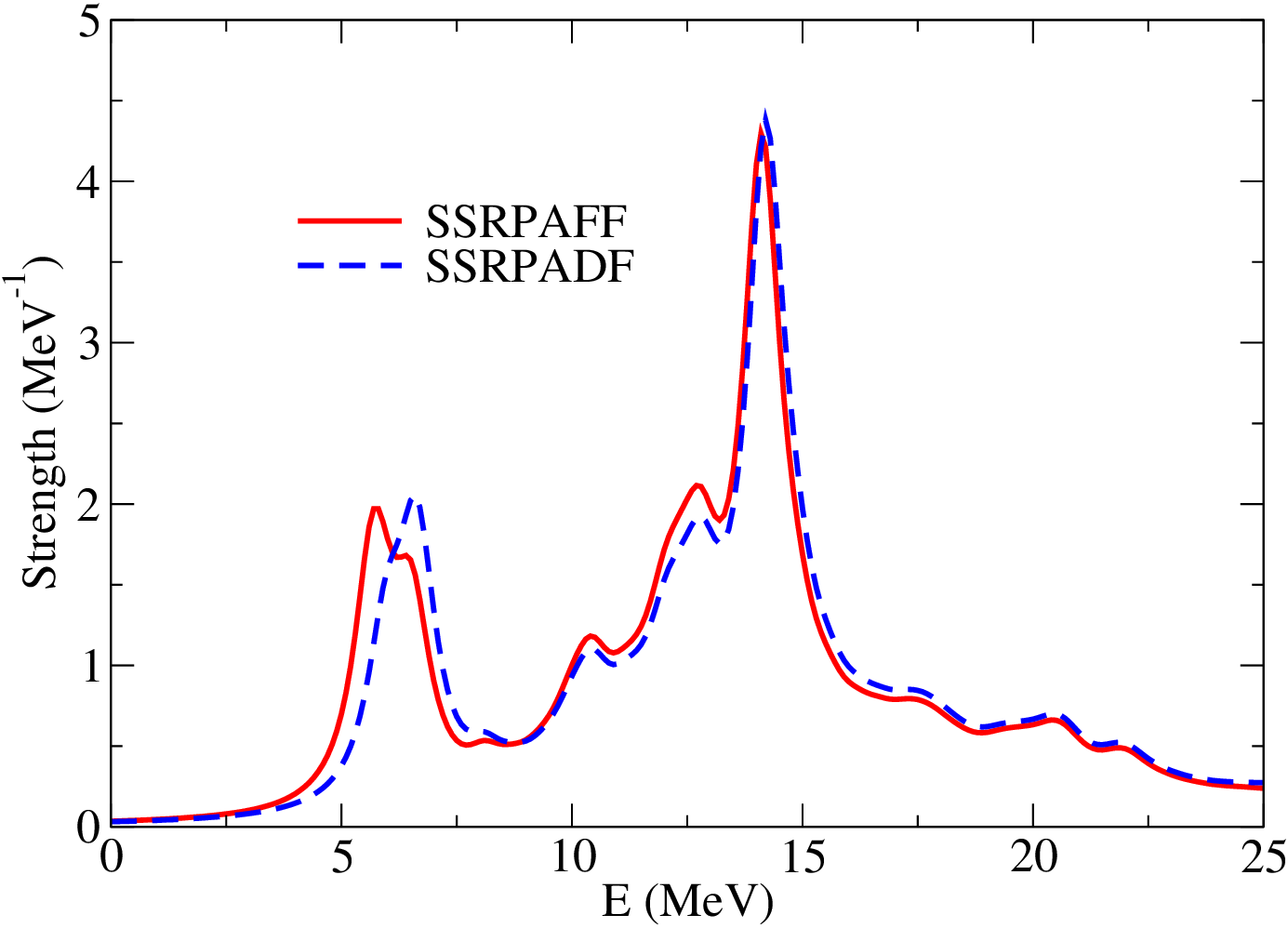}
\caption{GT$^-$ strengths obtained with the Skyrme interaction SGII for the nucleus $^{90}$Zr in MeV$^{-1}$. SSRPAFF and SSRPADF spectra are shown. }
\label{diago}
\end{figure}

We show in Fig. \ref{diago} the comparison between the SSRPADF and SSRPAFF GT$^-$ spectra in $^{90}$Zr, having folded the discrete spectra with a 1-MeV-width Lorentzian. Differently from the case of lighter nuclei, the hybrid approximation, where the matrix to invert is treated as diagonal, seems to work very well in this case. The two spectra are indeed very similar and almost superposed, especially at higher energies. We may conclude that, when the number of 2p2h configurations becomes very large in nuclei located beyond the medium-mass region (the number of configurations becomes typically larger that  $5 \cdot 10^4$) the off-diagonal terms may be safely neglected in the matrix to invert without loosing any important information in the excitation spectrum. We have indeed checked that this is the case by using different kind of Skyrme forces. In the present section, we will thus present  SSRPADF calculations for the two heavier nuclei that we treat.  

\begin{figure}
\includegraphics[scale=0.32]{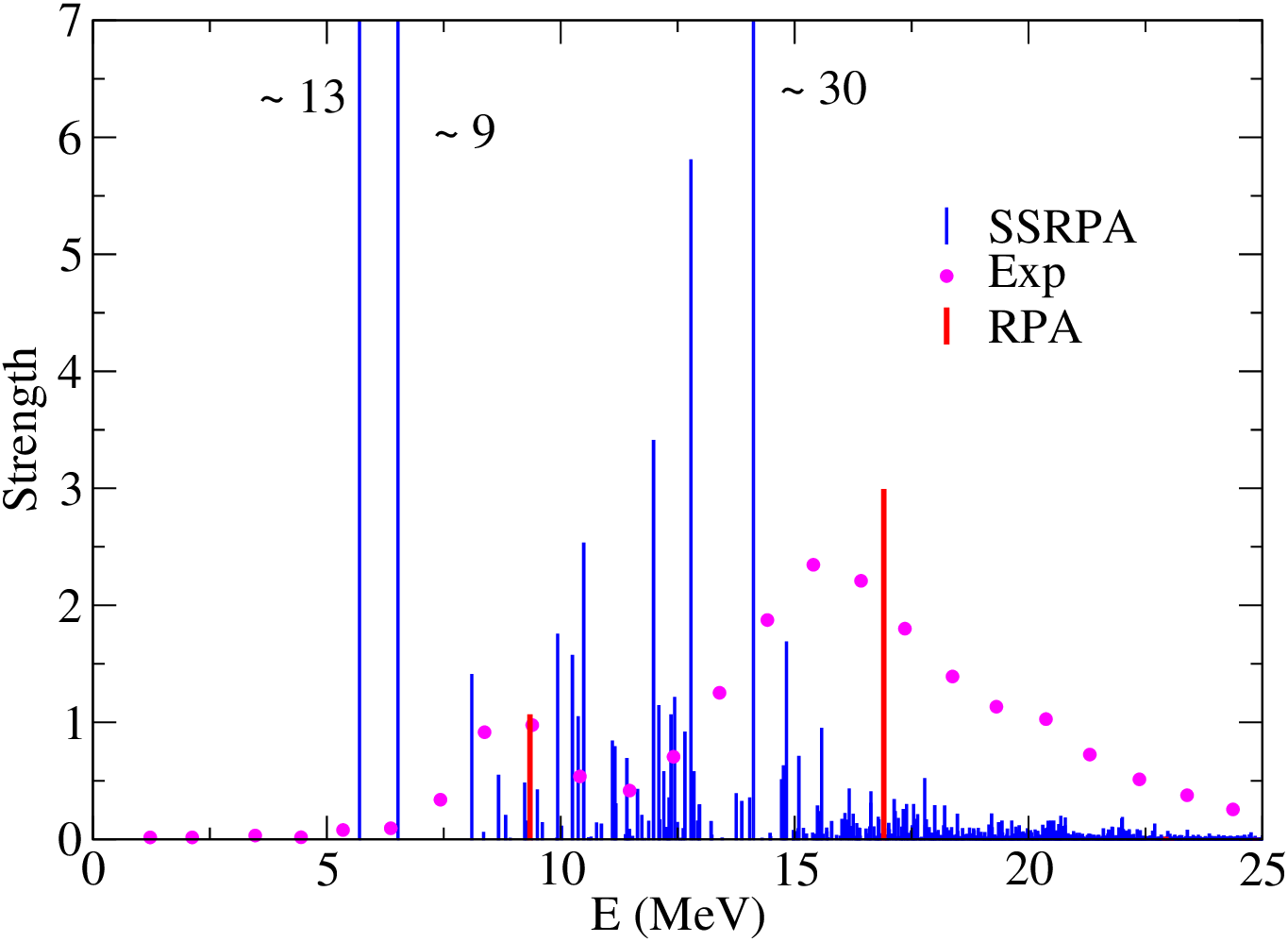}
\caption{GT$^-$ SSRPA and RPA discrete spectra obtained with the Skyrme interaction SGII for the nucleus $^{90}$Zr (no units). The RPA discrete spectrum is divided by 7.3 and the SSRPA one is multiplied by 5. 
The experimental results, in MeV$^{-1}$, are extracted from Ref. \cite{zr2}. }
\label{fineZr}
\end{figure}

To see how the fine structure of the excitation spectrum is described, we show in Fig. \ref{fineZr} the SSRPA and RPA GT$^{-}$ discrete spectra (no units) and the experimental spectrum of Ref. \cite{zr2} (in MeV$^{-1}$). 
The theoretical spectra are not folded to avoid any widths and fragmentations artificially induced by the folding itself, especially in the RPA case, where the number of discrete peaks is very low.  
The absolute values of the theoretical bars and the experimental strengths cannot of course be compared to each other since they are expressed in different units, whereas it is meaningful to compare the positions of the peaks and the fragmentation of the strength.
To better visualize all the results in the same plot, the discrete spectra have been scaled, the RPA one being divided by 7.3 and the SSRPA one being multiplied by 5. 
 We immediately observe that, whereas the RPA spectrum displays only two discrete main peaks, the SSRPA one has a quite dense distribution of peaks, describing much more realistically the experimental distribution of the strength. We see that, with the used interaction SGII, the SSRPA spectrum is slightly shifted to lower energies compared to the experimental one whereas the RPA one slightly overestimates the excitation energies. 

\begin{figure}
\includegraphics[scale=0.32]{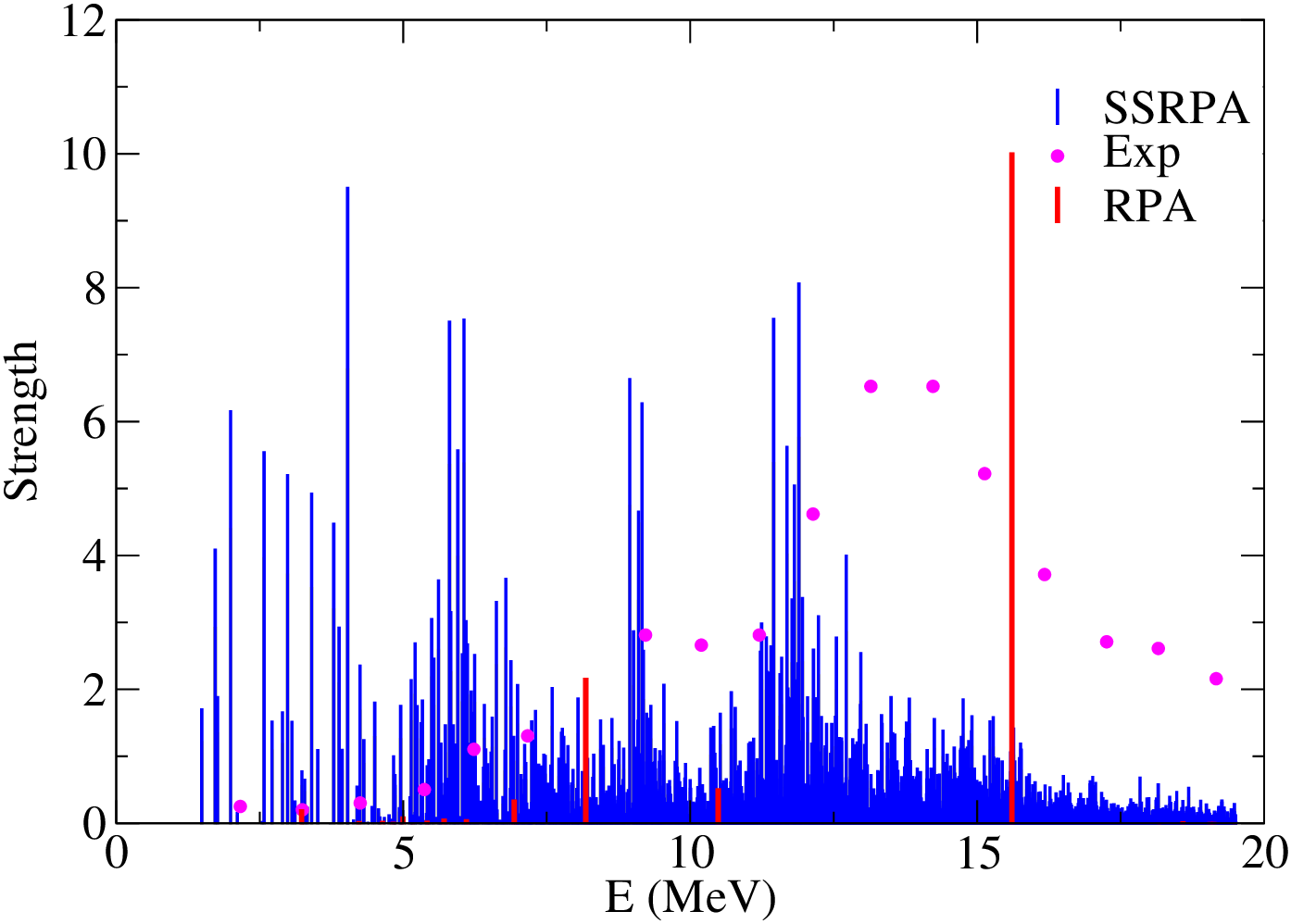}
\caption{GT$^-$ SSRPA and RPA discrete spectra obtained with the Skyrme interaction SGII for the nucleus $^{132}$Sn (no units). The RPA spectrum is divided by 7 and the SSRPA one is multiplied by 14. The experimental results, in MeV$^{-1}$, are extracted from Ref. \cite{sn}. }
\label{fineSn}
\end{figure}

The same analysis can be done for the heavier nucleus $^{132}$Sn. The experimental  (p,n) reaction was carried out for this nucleus at the Radioactive Isotope Beam Factory in RIKEN and the results were published in Ref. \cite{sn}. Also for this nucleus we compare, in Fig. \ref{fineSn}, the discrete RPA and SSRPA spectra (no units) with the corresponding experimental distibution (in MeV$^{-1}$).  
Again, since the absolute values of the theoretical and experimental strengths are not comparable, the RPA discrete spectrum has been divided by 7 and the SSRPA one has been multiplied by 14 to better visualize all the results in the same plot. 
The RPA spectrum has much less discrete peaks than the SSRPA one which provides naturally a fragmented and more dense response distribution.  Again, the SSRPA spectrum is slightly shifted towards lower energies compared to the experimental one whereas the RPA one slightly overestimates the excitation energies. 

The fact that we have found, for the two nuclei, an overall small shift of the spectra to lower energies compared to experimental results is something that depends on the choice of the effective interaction. We may in fact expect that other Skyrme interactions would provide spectra slightly shifted towards higher or lower energies. For example, it is known that a class of Skyrme interactions, called SAMi, was introduced in Ref. \cite{roca} and tailored specifically for describing spin-isospin excitations within the RPA model (and only within this model). 

\begin{figure}
\includegraphics[scale=0.32]{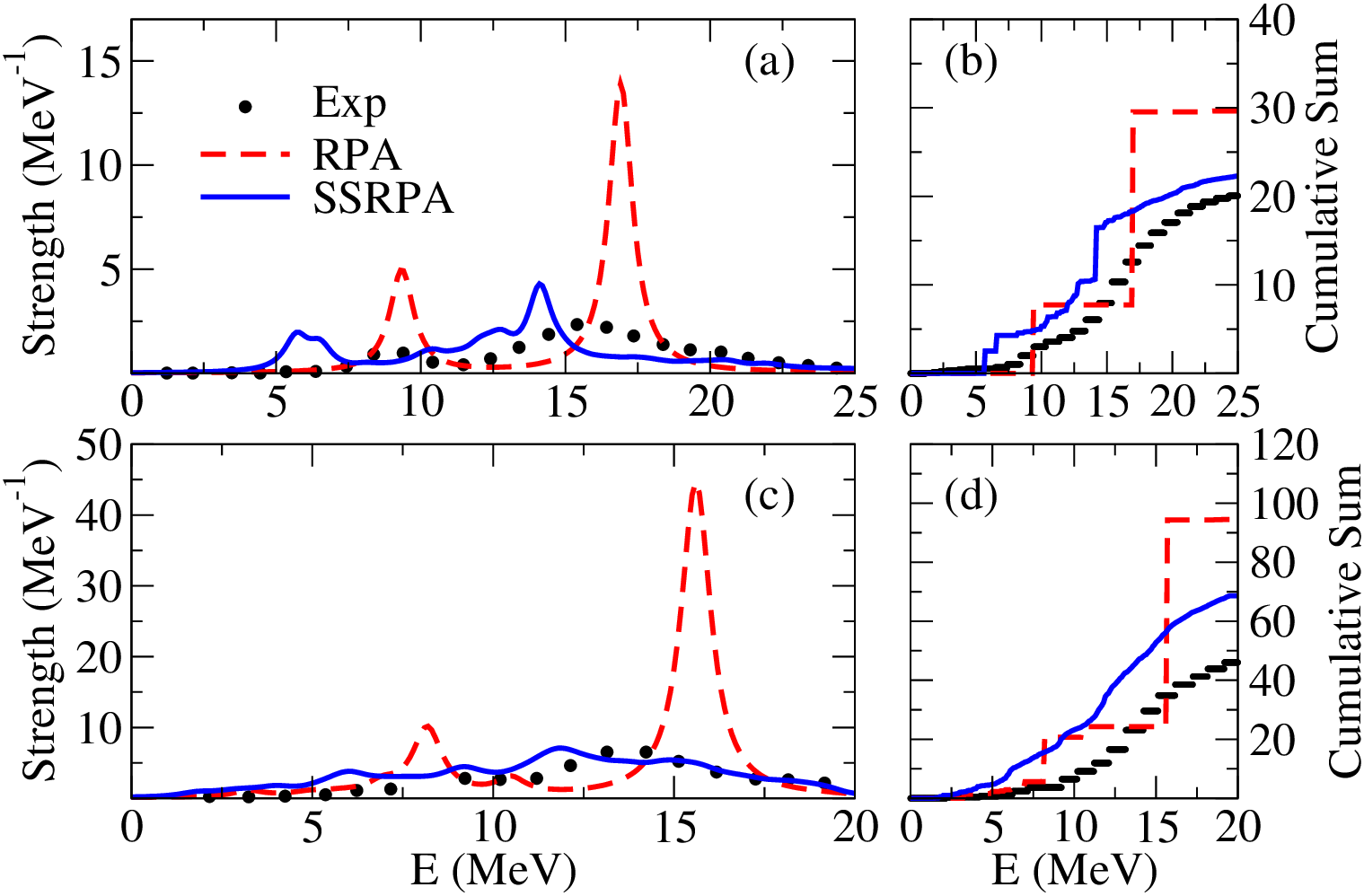}
\caption{(a) GT$^-$ SSRPA and RPA folded spectra obtained with the Skyrme interaction SGII for the nucleus $^{90}$Zr. 
The experimental results Ref. \cite{zr2} are also shown; (b) Strenghts of panel (a) integrated up to 25 MeV; (c) Same as in (a) but for the nucleus $^{132}$Sn. The experimental results are extracted from Ref. \cite{sn}; (d)  Strenghts of panel (c) integrated up to 20 MeV. }
\label{quen}
\end{figure}

Whereas the fragmentation will be described in a similar way by many different Skyrme interactions (this represents a genuine SSRPA effect), the specific position of the peaks in the spectrum will be slightly interaction-dependent.
As a matter of fact, another genuine SSRPA effect can be identified in the integrated strengths and, in particular, in their quenching. 
 In this respect, we show in Fig. \ref{quen} the folded SSRPA and RPA strengths (the folding is done with a 1-MeV-width Lorentzian) and the experimental strengths for $^{90}$Zr (a) and $^{132}$Sn (c), together with the cumulative sums of the strengths, evaluated up to 25 MeV for $^{90}$Zr (b) and up to 20 MeV for $^{132}$Sn (d). 

For the nucleus $^{90}$Zr, calculations carried out by Drozdz et al. \cite{dro} (within the SRPA model) and by Bertsch and Hamamoto \cite{bertsch} (within a perturbative approach) were reported in Ref. \cite{zr1}. However, those old SRPA calculations could be carried out, at that time, only by resorting to strong cuts and approximations.  It is known that, during the last years, this theoretical model was strongly improved and refined both in its formal aspects and in its numerical implementations, for example by ourselves \cite{gamba2020,gamba2015}. 
We thus cannot regard anymore those old results as robust references to which compare our present predictions.

A modern beyond-mean-field computation for the GT response in $^{90}$Zr was recently reported in Ref. \cite{robin}. These calculations were carried out within an approach based on the relativistic particle-vibration coupling and, for the first time, ground state correlations were included in the parent nucleus, coming from the particle-vibration coupling itself. A good agreement with the experimental strength distribution was found by the authors and commented in the article. However, the cumulative sum of the strength was not computed and shown there (to which we could compare our integrated strength). The authors of Ref. \cite{robin} mentioned by the way in this respect that the inclusion of the new correlations in the ground state of the parent nucleus induces a small violation of the Ikeda sum rule, as was already discussed in Ref. \cite{tse2007}. 

The only integrated strength we can refer to for this nucleus is the one coming from RPA calculations, which is shown in Fig. \ref{quen} for SGII. Other RPA strength distributions  and cumulative sums were also shown for $^{90}$Zr for example in Ref. \cite{cao}, obtained by using different Skyrme interactions. However, independently of the interaction, the RPA strength integrated up to 25 MeV is in all cases already almost equal to 30, much larger compared to the experimental value (b). Panel (b) shows that the SSRPA computation leads to a strongly improved agreement with the experimental integrated strength, providing a much lower value than in RPA (22.3) at the excitation energy of 25 MeV. The corresponding experimental value is slightly above 20, as one can see in the figure. 
The slope of the SSRPA curve nicely follows the experimental one and the sligth shift which is visible is generated by the already mentioned small shift (to lower values) in the excitation energies compared to the experiment.

For the nucleus $^{132}$Sn, modern beyond-mean-field particle-vibration-coupling calculations are available in the literature, within both non-relativistic \cite{niu} and relativistic \cite{robin2} approaches. However, only the authors of Ref. \cite{niu} presented in their article integrated strengths up to 20 MeV of excitation energy .  
Their results depend on the choice of a  smearing parameter $\Delta$. By increasing it, they can reduce the value of the cumulative sum. With the largest used value, $\Delta=1$ MeV, they obtain their best result, that is an integrated strength of 80 (at 20 MeV). Even if our prediction at 20 MeV  is located above the experimental value (we notice that the overall shift of the SSRPA spectrum to lower energies is also responsible for this) we may observe that we obtain anyway a better result than the one of Ref. \cite{niu}, 10 units lower than the particle-vibration coupling prediction. 

We may thus conclude  that, in general, the CE-SSRPA model seems to be more effective than other available beyond-mean-field EDF approaches in describing the quenching of GT strengths, globally leading to a much better agreement with the experimental measurements.

\section{Conclusions} \label{conclu}

We have applied the CE-SSRPA model, recently introduced in Ref. \cite{gamba2020} for $^{48}$Ca and $^{78}$Ni, to other closed-shell and closed-subshell nuclei located in different regions of the nuclear chart.  

A preliminary analysis is carried out in $^{48}$Ca to check the validity of different approximations. This study reveals that, for this nucleus (and for the lighter ones that we have treated here) the inversion of the $A_{22}$ matrix in the subtraction procedure must be performed without any type of approximations. If in particular the matrix to invert is approximated as diagonal the quality of the predictions results deteriorated, although a considerable improvement with respect to RPA is still obtained. 

Two lighter nuclei are analyzed, $^{14}$C and $^{22}$O to compare the SSRPA cumulative sums of the strength with the {\it{ab-initio}} coupled-cluster results of Ref. \cite{ekstrom}. This comparison indicates that, differently from the case of Ref. \cite{ekstrom}, the inclusion of a two-body part in the GT transition operator does not seem to be a crucial ingredient within our model to account for the quenching of the GT Ikeda sum rule. 

Finally, two heavier nuclei are investigated, $^{90}$Zr and $^{132}$Sn. We have checked that, for these heavier nuclei, where the number of 2p2h configurations considerably increases, the matrix to invert in the subtraction procedure can be simplified and safely taken as diagonal, without inducing important changes in the spectra. 
 The analysis of the strengths integrated up to excitation energies between 20 and 25 MeV reveals that, as was already found for $^{48}$Ca, the CE-SSRPA model  incorporates the necessary correlations to provide a much more effective description of the quenching of GT spectra and, consequently,  a much better agreement with experimental data.

%

%
%
%

\end{document}